\documentclass[12pt,nohyper,notoc]{JHEP}
\input epsf
\epsfverbosetrue
\usepackage{amsmath,euscript,array,amssymb,cite} 
\def\Dbarslash{\,\,{\raise.15ex\hbox{/}\mkern-12mu {\bar\D}}}
\def\Dslash{\,\,{\raise.15ex\hbox{/}\mkern-12mu \D}}
\def\delslash{\,\,{\raise.15ex\hbox{/}\mkern-9mu \partial}}
\def\delbarslash{\,\,{\raise.15ex\hbox{/}\mkern-9mu {\bar\partial}}}

\def\Z{{\EuScript Z}}

\def\P{{\cal P}}

\def\cl{{\,\rm cl}}
\def\lambdabar{\bar\lambda}
\def\R{{R}}
\def\psibar{\bar\psi}

\def\infinity{\infty}

\def\zero{{\scriptscriptstyle(0)}}
\def\new{{\scriptscriptstyle\rm new}}

\def\uA{\,\lower 1.2ex\hbox{$\sim$}\mkern-13.5mu A}
\def\uX{\,\lower 1.2ex\hbox{$\sim$}\mkern-13.5mu X}
\def\uD{\,\lower 1.2ex\hbox{$\sim$}\mkern-13.5mu {\rm D}}

\def\uF{\,\lower 1.2ex\hbox{$\sim$}\mkern-13.5mu F}
\def\uW{\,\lower 1.2ex\hbox{$\sim$}\mkern-13.5mu W}
\def\uWbar{\,\lower 1.2ex\hbox{$\sim$}\mkern-13.5mu {\overline W}}
\def\uV{\,\lower 1.2ex\hbox{$\sim$}\mkern-13.5mu V}
\def\uv{\,\lower 1.0ex\hbox{$\scriptstyle\sim$}\mkern-11.0mu v}
\def\uPsi{\,\lower 1.2ex\hbox{$\sim$}\mkern-13.5mu \Psi}
\def\uPhi{\,\lower 1.2ex\hbox{$\sim$}\mkern-13.5mu \Phi}
\def\uchi{\,\lower 1.5ex\hbox{$\sim$}\mkern-13.5mu \chi}
\def\Psibar{\bar\Psi}
\def\uPsibar{\,\lower 1.2ex\hbox{$\sim$}\mkern-13.5mu \Psibar}
\def\upsi{\,\lower 1.5ex\hbox{$\sim$}\mkern-13.5mu \psi}
\def\psibar{\bar\psi}
\def\upsibar{\,\lower 1.5ex\hbox{$\sim$}\mkern-13.5mu \psibar}
\def\upsibarzero{\,\lower 1.5ex\hbox{$\sim$}\mkern-13.5mu \psibar^\zero}
\def\ulambda{\,\lower 1.2ex\hbox{$\sim$}\mkern-13.5mu \lambda}
\def\ulambdabar{\,\lower 1.2ex\hbox{$\sim$}\mkern-13.5mu \lambdabar}
\def\ulambdabarzero{\,\lower 1.2ex\hbox{$\sim$}\mkern-13.5mu \lambdabar^\zero}
\def\ulambdabarnew{\,\lower 1.2ex\hbox{$\sim$}\mkern-13.5mu \lambdabar^\new}
\def\D{{\cal D}}

\def\Dslash{\,\,{\raise.15ex\hbox{/}\mkern-12mu \D}}
\def\Dbarslash{\,\,{\raise.15ex\hbox{/}\mkern-12mu {\bar\D}}}
\def\delslash{\,\,{\raise.15ex\hbox{/}\mkern-9mu \partial}}
\def\delbarslash{\,\,{\raise.15ex\hbox{/}\mkern-9mu {\bar\partial}}}
\def\uAcl{\,\lower 1.2ex\hbox{$\sim$}\mkern-13.5mu A^{}_{\cl}}
\def\uAbarcl{\,\lower 1.2ex\hbox{$\sim$}\mkern-13.5mu A_{\cl}^\dagger}

\def\uA{\,\lower 1.2ex\hbox{$\sim$}\mkern-13.5mu A}

\def\zero{{\scriptscriptstyle(0)}}

\def\beqa{\begin{eqnarray}} 
\def\eeqa{\end{eqnarray}} 
\def\beq{\begin{equation}} 
\def\eeq{\end{equation}} 
\def\R{\mbox{\rm I\kern-.18em R}} 
 
\def\P{\mbox{\rm I\kern-.18em P}} 
\def\uno{\mbox{1 \kern-.59em {\rm l}}} 
\def\Z{{Z \kern-.45em Z}} 
\def\Q{{\kern .1em {\raise .47ex \hbox{$\scriptscriptstyle |$}} 
\kern -.35em {\rm Q}}}

\def\cl{\mbox{\scriptstyle cl}}

\font\mybb=msbm10 at 12pt

\def\bb#1{\hbox{\mybb#1}} 
\def\Z {\bb{Z}}

\newcommand{\n}{\nonumber}

\renewcommand{\l}{\Lambda}
\renewcommand{\a}[1]{\alpha_{#1}}
\usepackage{graphicx}

\font\tenbbb=msbm12 \font\sevenbbb=msbm7         
  
\newfam\bbbfam \newfam\amsfam \newfam\bmfam      
\textfont\bbbfam=\tenbbb \scriptfont\bbbfam=\sevenbbb
\def\dbl {\fam\bbbfam}    

\title{On The Strong-Coupling Spectrum of Pure SU(3) Seiberg-Witten Theory}

\author{Brett J. Taylor \\
Department of Physics, University of Wales Swansea,
Swansea, SA2 8PP, UK\\
E-mail: {\tt pybr@swansea.ac.uk}}

\abstract{We consider the two complex dimensional moduli space of supersymmetric vacua 
for low energy effective ${\cal N}=2$ SYM with gauge group SU(3). 
We describe, at the topological level, a consistent model of
how the relevant curves of marginal stability (CMS) intertwine with the
branch cuts to partition the moduli space into pieces carrying different BPS 
spectra. 
At strong coupling we find connected {\em cores} which carry
a smaller BPS spectrum than that at weak coupling. At the strongest coupling
we find {\em double cores} which carry a finite BPS spectrum. These include not 
only states one can deduce from the monodromy group, but three states, bounded away from weak coupling, 
each of which we interpret as a bound state of two BPS gauge bosons.
We find new BPS states at weak coupling corresponding to a excitations of a state with magnetic charge a simple co-root, with respect to the other simple root direction.} 

\preprint{{\tt hep-th/0107016} \\ SWAT/311}


\begin{document}

\subsection{Introduction}

The pioneering work of Seiberg and Witten \cite{sw1,sw2} gave exact non-perturbative results
for Wilsonian low energy effective ${\cal N}=2$ supersymmetric gauge theory in
the case of gauge group SU(2) and with 0, 1, 2, 3 and 4 hypermultiplets in the
fundamental representation. Many hints were given as to how this could be 
extended to higher gauge groups or in other directions. These were quickly 
seized upon and fleshed out in some detail by many groups 
\cite{kly,af,kl1,klt,hanoz,bl,dsund,as,dsund2,hiha}.

Generic to all the SU(2) models seemed to be the existence of a {\em core} region
at strong coupling, bounded from the weakly coupled region by a curve of 
marginal stability (CMS) whereupon hitherto stable BPS states become marginally
stable and decay into two (lighter) elements of a reduced spectrum which lies 
within the core.

Seiberg and Witten did not pursue the calculation to find the position of the 
CMS, 
leaving us to wait for the rigorous analysis of Bilal and Ferrari \cite{bf,bfb,bfc}. Rather, 
they contented themselves with an argument for its existence based on the
consistency of the weakly coupled BPS spectrum. Only three states in this 
spectrum transform through the quantum branch cut back into another state
of the spectrum (another one of the three). They argue that a CMS must be 
present surrounding this branch cut whereupon all the other states in the 
weak-coupling spectrum decay to the favoured three.

In this paper we shall attempt a similar argument for the SU(3) moduli space.
We first consider the case of weak coupling. 
We then present a consistent example of how this may be continued to when 
an SU(2) subalgebra corresponding to a root direction becomes strongly coupled,
and finally to when two such subalgebras have strong coupling. 
 
An initial inspection would conclude that the BPS spectrum at weak coupling 
consists not of just one tower of dyons, but of 
three. There are many additional complications, one of which, 
described in \cite{timchrist} by Hollowood and Fraser, 
being that there are disconnected regions of moduli space at weak coupling 
carrying different towers of BPS 
dyons, separated at both weak and strong coupling by a combination of branch cuts and curves of marginal stability.
However, for gauge groups of rank greater than one, such as in this case, 
the two Higgs fields are generically non-aligned in group space. It has been 
shown \cite{jerome2} that in this instance many 
more BPS towers exist at weak coupling, indeed a countably infinite number. 
All except the previously mentioned three of these exist as higher spin 
multiplets, and are labelled by an integer $m$ which is twice the spin 
of the highest spin state in the multiplet.
We find that as we increase the strength of the coupling these decay on curves of marginal stability into smaller 
multiplets until we reach a region where we are left with the three tower 
situation found in \cite{timchrist, tim1, tim2}.

We shall find curves which exist only at strong coupling, as well as ones which
stretch to and can be extrapolated from weak coupling.
When only one of the sub-SU(2)'s is strong, there is a three-dimensional bounded region, 
called the core, containing a reduced BPS spectrum analogous to the SU(2) case.
Within 
self-intersections of this space lie disconnected bounded regions with a 
further reduced spectrum (double cores) which correspond to the case when the
sub-SU(2)'s of two roots are strongly coupled. 
Created at the double cores, but spreading throughout the strongly-coupled 
region we find the existence of three states $P_{i}$ with no magnetic charge, 
and electric charge equal to the difference of two roots (counting $-(\a{1}+\a{2})$ as the third root). We interpret these as bound states of massive gauge 
bosons.

This model predicts the existence of more new towers at weak coupling which we 
believe to have been overlooked. These can be created by passing 
already discovered higher spin multiplets through branch cuts with classical 
monodromies. This creates states which, like the ones found in 
\cite{jerome2}, have electric charges which together fill out half the root 
lattice at any particular point, but have simple co-roots as magnetic charge.
We are thus led to speculate that dyons with simple co-root magnetic charges 
at weak coupling, like those with non-simple co-roots, have low energy dynamics
determined by supersymmetric quantum mechanics on a hyperk\"{a}hler space of 
more than four real dimensions.
Exciting momenta in one of the new directions corresponds to a sort of Witten 
effect involving a charge non-proportional to the magnetic one.  
This is somehow restricted to the more usual subspace 
${\dbl R}^{3}\times S^{1}$, most probably due to the potential from a 
triholomorphic Killing vector field on the relative
moduli space blowing up when the two Higgs fields happen to align. 

Interestingly the mechanism can be seen explicitly at strong coupling, 
where states with these charges are created when one repeatedly wraps the normal 
simple root states about the core where the other simple 
root direction is strongly coupled. More than that, the charge picked up upon 
each 
rotation is that of the gauge boson bound states $P_{i}$, which exist at, and 
only at, strong coupling. In this core it seems one is allowed to excite
in this novel direction, but curves of marginal stability restrict the 
spectrum to only zero and one units in the proportional direction, 
as for the SU(2) case.

We shall choose to slice up our 
four-dimensional space into concentric three spheres as suggested by Argyres 
and Douglas \cite{ad1}, and to study the picture at various generic radii. 
We begin with large radius which corresponds to weak coupling, and then steadily
decrease the radius until we have mapped out the entire space.

\subsection{The Perturbative Weak-Coupling BPS Spectra}

Seiberg-Witten theory is a Wilsonian low-energy effective theory that is 
perturbatively 1-loop exact, and non-perturbatively affected by a series of instanton corrections. In ${\cal N}=1$ language
\[ {\cal L}_{eff} = \frac{1}{4\pi}{\rm Im}\left[ \int{d^{4}\theta
\frac{\partial {\cal F}(A)}{\partial A}\bar{A}}
+
\int{d^{2}\theta \frac{1}{2}\frac{\partial^{2}{\cal F}(A)}{\partial A^{2}}W^{\alpha}W_{\alpha}}\right] , \]
where ${\cal F}$ is the holomorphic prepotential. The ${\cal N}=2$ vector 
supermultiplet has two ${\cal N}=1$ components, $A$ being the chiral one, whose lowest component is $a$, and $V$ the vector one. $W_{\alpha}=-\frac{1}{4}\bar{D}^{2}D_{\alpha}V$ is the abelian 
field strength supermultiplet. We can split ${\cal F}$ into its constituent pieces
\[ {\cal F}= {\cal F}_{1-loop} + \sum_{n=1}^{\infinity}{{\cal F}_{n}\left(
\frac{\Lambda}{a}\right)^{\kappa n}}. \]  
At weak coupling the instantons are exponentially suppressed so we 
may concentrate on the perturbative analysis.

\FIGURE[h]{
\centering
\includegraphics[width=6cm]{csa1.pstex}
\put(-165,112){$\a{2}$}
\put(-90,155){$\a{3}$}
\put(-17,112){$\a{1}$}
\caption{Using $a$ as a coordinate the moduli space lives in the complexified Cartan subalgebra of SU(3). Shaded here are the weakly coupled regions, those bounded away from the weight directions. They extend off to infinity.} 
\label{csa1}}

If we find  $a_{D}=\partial {\cal F}/\partial a$ then it is a simple matter
to integrate to find ${\cal F}$.
For SU($N$) broken to U($1)^{N-1}$ one finds a sum over positive roots: 
\[ a_{D} = \frac{i}{2\pi}\sum_{\a{i}\in\Phi^{+}}{\a{i} (\a{i}.a)\left[1 +
\ln{\left(\frac{\a{i}.a}{\l}\right)^{2}}\right]}. \label{pertad} \]
One can read from this that the weak-coupling limit occurs when $|a.\a{i}| \gg 0$ for all the roots $\a{i}$. 
This is essentially the large $a$ limit, but excluding points close to
fundamental weight directions.
When $|a.\a{i}| = 0$ the gauge group is not generically broken, a non-abelian
subgroup reappears, the extra massless particles alter the effective theory, and this is realized as a singularity where the previous theory breaks down. 
For SU(3) the weakly coupled regions are sketched in fig.\ \hspace{-3pt}(\ref{csa1}). We label $\a{3}$ as lying between the other two roots (or $-\a{2}$ 
between $-\a{3}$ and $\a{1}$, or $-\a{1}$ between $-\a{3}$ and $\a{2}$).

In \cite{timchrist} Hollowood and Fraser consider the spectrum of BPS states
for the weakly coupled region with the two Higgs fields aligned in group 
space. For points which lie within the fundamental Weyl chamber 
chosen such that $\a{1}$ and $\a{2}$ are the simple roots, they find that the 
weakly-coupled space is split into two domains, each with a slightly different spectrum.
Corresponding to each of the two simple root directions there exists a tower of
dyons with magnetic charge equal to the co-root in question, and electrically 
charged with an integer multiple of the same root.
\[ Q_{1}^{n}= (\a{1},(n-1)\a{1}) \hspace{5pt};\hspace{15pt} 
Q_{2}^{n}= (\a{2},n\a{2}). \]

For a non-simple root it was shown in \cite{jerome2} using analysis in 
\cite{pope} that nowhere at weak coupling does a similar situation occur for 
$\a{3}$ with electric charge proportional to magnetic charge. Rather, the 
electric charge, which still remains in the root lattice, is shifted, equally 
for every state in the tower, in the direction of one of the other roots.
There are two possiblilities:
\[ Q_{3-}^{n}= (\a{3}, -\a{1}+n\a{3})\hspace{5pt};\hspace{15pt}
 Q_{3+}^{n}= (\a{3}, +\a{1}+n\a{3}). \]
Between them lies the barrier of a curve of marginal stability (CMS) which 
partitions the space in two. Note that if a BPS state is present, so is its
antimatter partner with the opposite charges, and therefore we shall ignore
any overall minus signs in the charge vectors. 

We now choose to parametrize the moduli space in terms of the Casimirs 
\[ u=\langle {\rm tr}(\phi^{2})\rangle\hspace{5pt};\hspace{15pt}
v=\langle {\rm tr}(\phi^{3})\rangle. \]
Following Argyres and Douglas \cite{ad1}, we foliate the space by 
hypersurfaces with the topology of   
three-spheres, 
\[ 4|u|^{3}+27|v|^{2}=R^{6}, \]
and study the three-space produced upon stereographic projection of each of these slices --- {\it i.e.}\ by the map
\[ ({\rm Re}\,u,{\rm Im}\,u,{\rm Re}\,v,{\rm Im}\,v) \rightarrow 
(\frac{{\rm Re}\,u}{R^{3}-\sqrt{27}{\rm Im}\,v},\frac{{\rm Im}\,u}{R^{3}-\sqrt{27}{\rm Im}\,v},\frac{{\rm Re}\,v}{R^{3}-\sqrt{27}{\rm Im}\,v}). \]

The three-spheres have been chosen such that the intersection of the 
surface of singularities with a large three sphere is a closed curve in the
form of a knot, the {\it trefoil} which wraps thrice around an auxilliary torus in one direction 
as it wraps twice around the other cycle.

\FIGURE[h]{
\centering
\includegraphics[width=5cm]{trefcl.pstex}
\caption{The classical trefoil ${\cal T}_{cl}$ wrapping three times around one cycle as it wraps twice around the other.}}

The large $R$ limit corresponds to the large $a$ limit, and so contains the 
weak-coupling limit when far from a singularity. In our picture this corresponds either to the point at infinity or at the origin. 
Together with gauge bosons $W_{1}$, $W_{2}$ and $W_{3}$, the set
$\{ Q_{1}^{n}, Q_{2}^{n}, Q_{3-}^{n} \} $ is the massive BPS spectrum around 
infinity,
and $\{ Q_{1}^{n}, Q_{2}^{n}, Q_{3+}^{n} \} $ that near the origin.

As can be seen from the perturbative formula for $a_{D}$, upon 
traversing a small loop around the singular surface, $a$ and $a_{D}$ mix.
This is an example of monodromy.
One can realize this by the strategic placement of branch cuts emanating from 
the singular surface. Associated to each cut is a matrix with integer entries
which performs the mixing between $a$ and $a_{D}$ (and any other global U(1)
charges present).
The mass of a BPS state with (magnetic, electric) charge $(g,q)$ is given by 
the modulus of a central charge $Z = g.a_{D}-q.a$, hence one can see that 
mondromy can also be realised by acting on the charges $(g,q)$ by the 
inverse of the monodromy matrix.
Following \cite{kl1} we choose to use the basis (Dynkin, fundamental) for our charges, in which $(\a{1},0)=(1,0,0,0)$, $(\a{2},0)=(0,1,0,0)$, $(0,\a{1})=(0,0,2,-1)$, and $(0,\a{2})=(0,0,-1,2)$. 
The monodromy matrices form a group, a subgroup of Sp($2(N-1),{\dbl Z}$).   

The branch cuts must be of real-codimension one, so will have
sections on our three-sphere that form a (self-intersecting) surface.
The only two choices consistent with the monodromy group are the ruled surfaces
formed as the locus swept out by lines connecting each point on the trefoil
to either a point near the origin, or to a point near infinity (or, of 
course, by adding any number of non-topology changing perturbations to this). 
We choose to pick our cuts all meeting at the origin. 
Our first choice is then to include the following three mondromy matrices 
given in \cite{kl1} as in fig.\ \hspace{-3pt}(\ref{tref1}).

\FIGURE[h]{
\centering
\makebox[6cm]{\includegraphics[width=6cm,height=5cm]{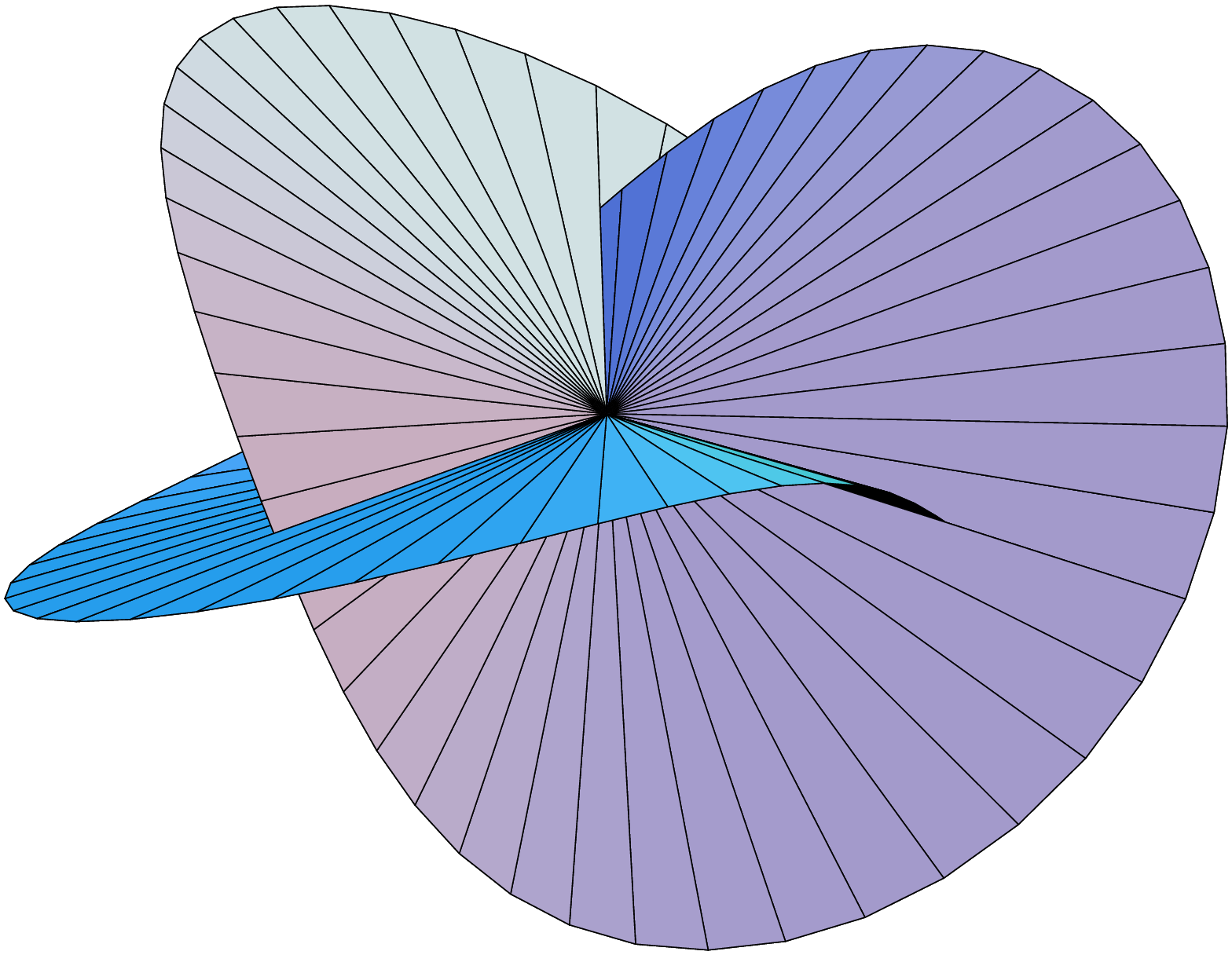}
\put(-115,110){$N_{1}$}
\put(-77,95){$N_{3-}$}
\put(-35,110){$N_{2}$}
\thicklines
\put(-110,70){\vector(0,1){35}}
\put(-72,45){\vector(0,1){45}}
\put(-30,70){\vector(0,1){35}}}
\caption{The classical branch cuts form a self-intersecting surface.}
\label{tref1}}

\[ N_{1}=
\left(
\begin{array}{cccc}
-1 & 0 & 4 & -2 \\
1 & 1 & -2 & 1 \\
0 & 0 & -1 & 1 \\
0  & 0 & 0 & 1
\end{array}
\right);
\hspace{20pt}
N_{2}=
\left(
\begin{array}{cccc}
1 & 1 & 1 & -2 \\
0 & -1 & -2 & 4 \\
0 & 0 & 1 & 0 \\
0  & 0 & 1 & -1
\end{array}
\right);
\]
\[ N_{3-}=
\left(
\begin{array}{cccc}
0 & -1 & 1 & -2 \\
-1 & 0 & 4 & 1 \\
0 & 0 & 0 & -1 \\
0  & 0 & -1 & 0
\end{array}
\right).
\]

\FIGURE[h]{
\makebox[6cm]{\hspace{-11cm}\includegraphics[width=3cm]{intersection.pstex}
\put(-90,70){$N_{2}$}
\put(-139,-10){$N_{3-}=N_{2}N_{1}N_{2}^{-1}$}
\put(-23,84){$N_{1}$}}
\caption{A typical intersection.}
\label{tref2}}

The classical monodromy group generated by any two of these linear maps has a 
${\dbl Z}_{3}$ symmetry upon cyclically permuting the indices 1, 2 and $3-$.
The relations are 
\[ N_{3-}=N_{1}^{-1}N_{2}N_{1}=N_{2}N_{1}N_{2}^{-1} \hspace{10pt} {\rm and}
\hspace{5pt}{\rm cyc. \hspace{10pt}perms}\label{relat}\]
One of these can be seen in fig.\ \hspace{-3pt}(\ref{tref2}), a close-up of an 
intersection in fig.\ \hspace{-3pt}(\ref{tref1}). The relation above 
also justifies the imposition of a common meeting point for the cuts --- all 
small closed loops in the vicinity of the origin have trivial monodromy.

The classical mondromy associated with each root acts on states with the same 
subscript by shifting them two places higher up their respective tower. 
It moves states
in the other towers to different towers, sometimes to a tower we have met 
already, 
and sometimes to more exotic ones, 
which we shall encounter later.
\[ Q_{i}^{n}N_{i}=Q_{i}^{n-2} \]
\begin{eqnarray}
 Q_{1}^{n+2}N_{2} = Q_{3+}^{n} & = & Q_{2}^{n}N_{1}^{-1}, \n\\
 Q_{2}^{n}N_{1} = Q_{3-}^{n} & = & Q_{1}^{n}N_{2}^{-1}.
\end{eqnarray}
It still remains to place the curve of marginal stability described in
\cite{timchrist}. If, as in fig.\ \hspace{-3pt}(\ref{radsec1}), 
we take 2-dimensional sections of our space by 
considering a vertical half-plane projecting radially outwards from $u=0$, 
then the
trefoil will intersect it in two points. 
We claim the CMS, defined as the set for which Im$[\a{1}.a/\a{2}.a]=0$, 
lies between them, and
for our purposes it will be sufficient to represent it as a straight line
between them.
This line will rotate as one rotates the 2-dimensional 
radial section about the $u=0$ axis, forming a twisted ribbon with the 
trefoil as its boundary.
\FIGURE[h]{
\centering
\makebox[6cm]{\includegraphics[width=6cm,height=5cm]{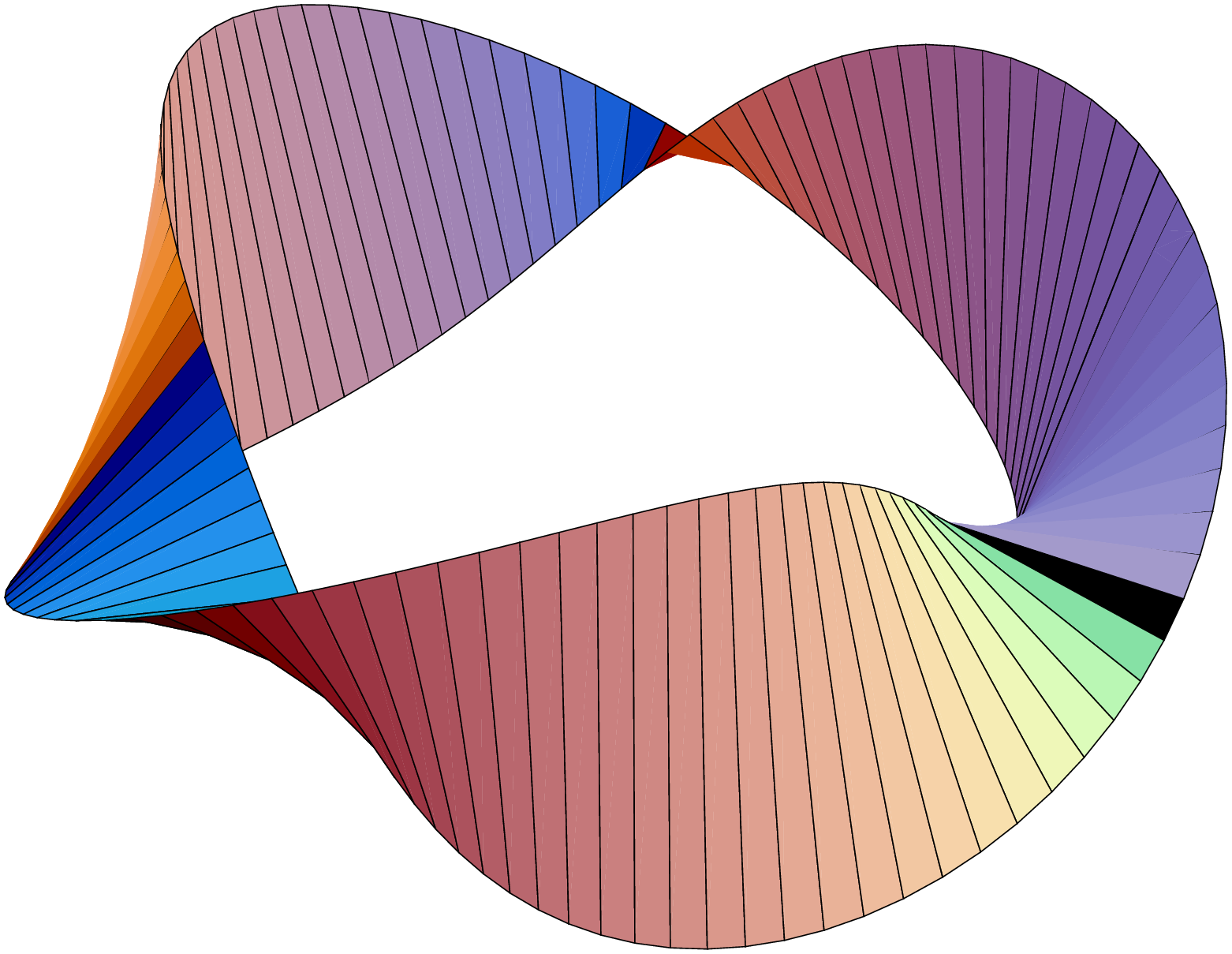}}
\caption{The classical CMS.}
\label{ribbon}}

\FIGURE[ht]{
\centering
\includegraphics[width=10cm]{radsec1.pstex}
\put(-46,96){$N_{1}$}
\put(-54,59){$N_{2}$}
\caption{Radial sections of curves aid exposition.}
\label{radsec1}}

From now on, we shall concentrate on the third of the space for which $N_{1}$
is the monodromy on top (with $N_{2}$ on the bottom) and call this the 
1,2 sector. Sometimes we choose
to show the branch cuts bent away from being straight to make space to 
describe all the states present between them.
Fig.\ \hspace{-3pt}(\ref{radsec2}) shows how the branch cuts and the CMS conspire to 
partition the 1,2 sector into two domains, one connected to the weak-coupling
limit at $u=0$ with $Q_{3+}$'s, not $Q_{3-}$'s in the BPS spectrum, the other to the one at 
$u=\infty$ with $Q_{3-}$'s, not $Q_{3+}$'s.

\FIGURE[ht]{
\makebox[10cm]{\includegraphics[width=8cm]{radsec2.pstex}
\put(-221,122){$N_{1}$}
\put(-221,-10){$N_{2}$}
\put(-186,126){$Q_{1}^{n}$}
\put(-142,126){$Q_{2}^{n}$}
\put(-97,126){$Q_{3+}^{n}$}
\put(-186,163){$Q_{1}^{n-2}$}
\put(-142,163){$Q_{3-}^{n}$}
\put(-97,163){$Q_{2}^{n}$}
\put(-186,-8){$Q_{1}^{n}$}
\put(-142,-8){$Q_{2}^{n}$}
\put(-97,-8){$Q_{3-}^{n}$}
\put(-186,29){$Q_{3+}^{n-2}$}
\put(-142,29){$Q_{2}^{n-2}$}
\put(-97,29){$Q_{1}^{n}$}
\put(-110,55){$Q_{3+}^{n}$}
\put(-136,101){$Q_{1}^{n}+Q_{2}^{n}$}
\put(2,55){$Q_{1}^{n+2}+Q_{2}^{n}$}
\put(2,101){$Q_{3-}^{n}$}}
\caption{The classical BPS spectra in the $3+$ and $3-$ domains.}
\label{radsec2}}

As the Casimir variables are invariant under the action of the Weyl group, we
have no right to expect that always the $\a{3}$ tower is deformed as opposed 
to the other two. 
Putting $\a{1}$, $\a{2}$ and $-\a{3}$ on the same footing, we 
allow for weak-coupling regions where any two of these roots are regarded as 
simple, while the tower corresponding to the third one has shifted electric 
charges. 

Therefore the space must be partitioned such as to have six disjoint 
weakly-coupled regions containing towers of BPS states $\{ 1,2,3-\}$,
$\{ 1, 2, 3+\} $, $\{ 1,2-,3\} $, $\{ 1,2+,3\} $, $\{ 1-,2,3\} $ and
$\{ 1+,2,3\} $. This can be performed by the addition of three branch cuts 
seen as vertical half-planes from $u=0$ to $u=\infty$, at angles such that
they project along the lines of intersection of the previously
described classical branch cut. The monodromies at these cuts are given by 
$UU_{i-}$ which are described by the following matrices in our basis, with directions as indicated in fig.\ \hspace{-3pt}(\ref{ucuts2}), where only the cuts and curves in the 1,2 sector are drawn.

\FIGURE[ht]{
\centering
\includegraphics[width=10cm]{ubitrib.pstex}
\put(-285,233){$UU_{1-}$}
\put(-70,257){$UU_{2-}$}
\put(-55,145){$UU_{3-}$}
\put(-250,200){\large $Q^{n}_{1}$ $Q^{n}_{2}$ $Q^{n}_{3-}$}
\put(-210,250){\large $Q^{n}_{2}$ $Q^{n}_{3}$ $Q^{n}_{1-}$}
\put(-50,210){\large $Q^{n}_{3}$ $Q^{n}_{1}$ $Q^{n}_{2-}$}
\caption{The U's rotate which of the towers of BPS states has shifted electric charge. Only one third of the curves is shown.}
\label{ucuts2}}

\[
U=
\left(
\begin{array}{cccc}
0 & 1 & 0 & 0 \\
-1 & -1 & 0 & 0 \\
0 & 0 & -1 & 1 \\
0 & 0 & -1 & 0
\end{array}
\right),
\hspace{20pt}
U_{1-}=
\left(
\begin{array}{cccc}
-1 & -1 & -2 & -1 \\
1 & 0 & 1 & 1 \\
0 & 0 & 0 & -1 \\
0 & 0 & 1 & -1
\end{array}
\right).\]

\[ U_{2-}=
\left(
\begin{array}{cccc}
-1 & -1 & 1 & 1 \\
1 & 0 & 1 & -2 \\
0 & 0 & 0 & -1 \\
0 & 0 & 1 & -1
\end{array}
\right),
\hspace{20pt}
U_{3-}=
\left(
\begin{array}{cccc}
-1 & -1 & 1 & -2 \\
1 & 0 & -2 & 1 \\
0 & 0 & 0 & -1 \\
0 & 0 & 1 & -1
\end{array}
\right).\]

The charges of the states not yet defined ({\it e.g.}\ $Q_{2+}^{n}$, $Q_{3}^{n}$), can be obtained by cyclically permuting $\a{1}$, $\a{2}$ and $-\a{3}$ from
the analogous state previously listed. Note $U_{3-}$ cyclically permutes 
$Q_{1}^{n}$, $Q_{2}^{n}$ and $Q_{3-}^{n}$, the BPS spectrum in the outer domain
of the 1,2 sector. Similarly for the other $U_{i-}$. There also exist 
$U_{i+}$'s permuting the towers of the inner spectra. 
$U$ permutes the subscripts 1, 2, and 3, acting as an exact ${\dbl Z}_{3}$
on the moduli space, rotating it about $u=0$ by $2\pi/3$.

\subsection{Higher Spin Multiplets}
Now let us consider the case for non-aligned Higgs fields. In \cite{jerome2}
it is claimed that the two BPS spectra for the 1,2 sector should contain BPS 
states in higher spin multiplets. The analysis, based on the Scherk-Schwarz mechanism,
results in the prediction of states with charges
$(\a{3},n\a{3}-s\a{1})$ for the outer domain, and $(\a{3},n\a{3}+s\a{1})$ for the inner domain, with $s\in{\dbl Z}^{+}$ being twice the 
spin of the highest state in the multiplet.
Passing the new outer states up through the classical branch cut, they are acted upon by the monodromy $N_{2}N_{1}$, giving states with charges $(\a{1}, (n-1)\a{1}+(s-1)\a{2})$. 
Repeating the circuit gives $(\a{2},n\a{2}+(s-1)\a{3})$, and a third repetition
 sees us return to the original higher spin towers. 
Therefore we predict the existence of these new towers of states at weak 
coupling. 
Similarly for the inner domain with the sign of $s$ reversed, and for the 2,3
and 3,1 sectors, with $\a{1}$, $\a{2}$ and $-\a{3}$ cyclically permuted.
Any branch cut or CMS which might posssibly alter the spectrum during the 
course of this triple circuit as a means of circumventing this existence 
argument would necessarily touch either $u=0$ or $u=\infty$
and so could only further partition the set of states at weak coupling, not
eliminate the space carrying the unwanted spectrum completely.  

It turns out that none of these higher spin states exist at strong coupling.
They all decay on a series of CMS's labelled $D$'s and $F$'s which circle 
around both arms (as viewed in the 2D section). We are therefore at liberty to 
consider both the quantum and classical pictures without the higher multiplets
 interfering.
We shall return to describe the necessary CMS's soon, after we have motivated 
the introduction of further notation.
  
\subsection{Non-Perturbative Effects for Weak and Moderate Coupling}

We now include the effect of instantons in this picture. For all SU($N$) these 
cause a bifurcation of the singular surface, 
seen as an effective shift in the top 
Casimir ($v$ for SU(3)) proportional to $\pm \l^{N}$.
For SU(2), a {\em quantum} branch cut appears between the two singular points,
and so one may trace paths braiding around the two singularities.
For SU(3), it is similarly the case that the two trefoils we now have will
possess a quantum branch cut between them. As Re[$v$] is the vertical direction
in the 3-dimensional sections we take, the trefoils are nothing but two
copies of the same trefoil vertically translated by an amount proportional to 
$\l^{3}$. The quantum branch cut is therefore a ribbon (with no
twists) composed of vertical lines between the two trefoils. 
(see fig.\ \hspace{-3pt}(\ref{treffy2})).

\FIGURE[h]{
\centering
\includegraphics[width=6cm]{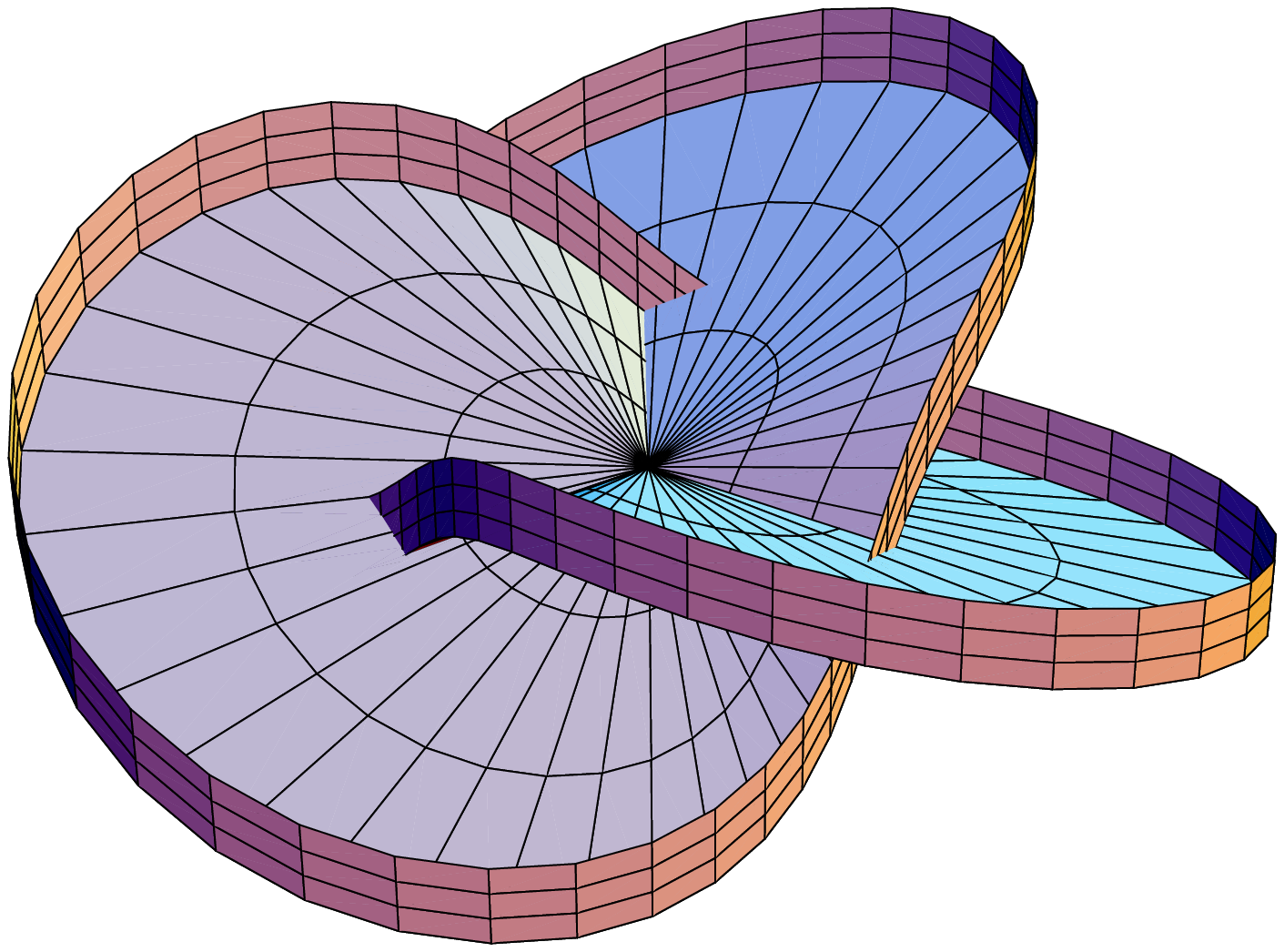}
\caption{The classical and quantum branch cuts for large $R$.}
\label{treffy2}}

The three classical monodromies split into six quantum ones, as detailed in 
\cite{kl1}. To each classical monodromy $N_{i}$ we associate a pair of
quantum monodromies $M_{2i}$ and $M_{2i-1}$, with $M_{2i}M_{2i-1}=N_{i}$.
As for SU(2), the $M$'s must be parabolic, meaning their repeated action
on a state adds a constant amount each time to the charges of that state, 
forming a tower isomorphic to ${\dbl Z}$. Thus we find for SU(3), acting in 
four real dimensions, that they each preserve a 2-dimensional subspace, acting as a 
shear on the whole charge space by a vector in the invariant subspace. 
Each $M$ is therefore specified uniquely
(modulo direction) by a charge vector $(g,q)$. By the mass formula, a BPS 
state with charge $(g,q)$ will become massless at a fixed point of this 
monodromy, {\it i.e.}\ at a singularity where the corresponding branch cut 
ends.

The matrix corrresponding to charge $(g,q)$ is given by
\[
M_{(g,q)}=
\left(
\begin{array}{c|c}
1+q\otimes g & q \otimes q \\ \hline
-g \otimes g & 1-g\otimes q
\end{array}
\right).
\]

We have some freedom to label charges, and we do so as to put ourselves on
a sheet where states have simple quantum numbers. 
Following \cite{kl1} we choose $M_{2}=M_{(1,0,0,0)}$, and 
$M_{3}=M_{(0,1,0,0)}$. 
The odd $M$'s, and similarly the even $M$'s obey the same relation as the 
$N$'s:
\begin{eqnarray} 
M_{6} & = & M_{2}^{-1}M_{4}M_{2}=M_{4}M_{2}M_{4}^{-1} \n\\
M_{5} & = & M_{1}^{-1}M_{3}M_{1}=M_{3}M_{1}M_{3}^{-1} \hspace{5pt}{\rm and}
\hspace{5pt}{\rm cyc.}
\end{eqnarray}
Due to the cyclic symmetry, because $M_{2}$ and $M_{3}$ commute (the charge vectors lie in each other's invariant subspace), $M_{4}$ and $M_{5}$ commute, as well as $M_{6}$ and $M_{1}$.
If we associate the even $M$'s with the cut continuing from their respective $N$, ($M_{2i}$ with $N_{i}$), pointing inwards toward the origin as in 
fig.\ \hspace{-3pt}(\ref{cuty2}), we can then find which states become massless at which places.
We always get the same pattern: 
For the arm with label $i\star$ (where $\star$ is blank, $+$ or $-$), the state becoming
massless at the top singularity is $Q_{i\star}^{1}$, and at the other is 
$Q_{i\star}^{0}$ from above and $Q_{i\star}^{2}$ from below. 
\FIGURE[h]{
\centering
\includegraphics[width=12cm]{goodtry.pstex}
\put(-292,0){$N_{2}$}
\put(-262,44){$N_{1}$}
\put(-84,-1){$N_{2}$}
\put(-32,12){$M_{4}$}
\put(-50,38){$N_{1}$}
\put(0,77){$M_{2}$}
\caption{Our choice of cuts.}
\label{cuty2}}

Thus, at this level, it seems each arm is just a little SU(2) microcosm, nothing more than an area where the SU(3) solution is dominated by an embedded SU(2) 
associated to one of the root directions. We shall continue the pursuit of this 
idea later.

In this region of moduli space we shall encounter two types of curves of 
marginal stability, the $L$'s and the $G$'s (for local and global). Further 
out, encircling both arms and never coming close to them are a collection of curves labelled $D$'s and $F$'s upon which all the higher spin states decay to the simpler spectrum given in \cite{timchrist}. 
We start with the premise that no CMS has a boundary, and that occasionally they may touch the singular surface. It is from this that we glean information about 
them. 
At any particular point on a CMS, a decay may take place involving one state
decaying into two others with charges that sum to the original. The curve
does not see these states in particular, but the whole lattice of charges 
generated by integer linear combinations of any two of them. Any possible decay
 involving three states with charges on the lattice which sum to zero may 
happen at some point along the curve, indeed many at once. Most will not
occur, either as none of the states is present, or for kinematic reasons,
or for some other reason that a decay into an exotic particle is excluded.
Sometimes, also, a decay occurs which has no effect on the spectrum as
parent and daughter states are present on both sides of the curve anyway.
If a curve has no effect on our spectrum at all, we say it is {\em irrelevant}.
If a decay takes place at one point of a curve, it will also occur in the 
vicinity of that point. It will continue along the curve, including through 
branch cuts (whereupon each state is individually acted upon by the monodromy) unless it 
reaches a point where the curve adjoins a singularity, whereupon the light 
particle `swaps sides', as which of the other two states is most massive changes, and the arrow of the decay process reverses. Arrow reversal is a necessary condition for the curve to touch a singularity.
Later we shall meet accumulation points, which are the next best thing to singular points, and the curves behave characteristically there too. 

We presume that if a curve passing by a singularity has upon it a decay involving the state which becomes massless at that singularity, then the curve will touch the singularity. As we rotate our 2-dimensional radial section, any
CMS (now a line) touching one of the four singularities will do so regardless
of the position of the singularity (the top and bottom arms, each with their singularity pair, will rotate around each other). 
Even when the singularity passes through a branch cut the curve can follow it, and the states involved in the decay will be the original ones acted upon by the monodromy (which is linear). We say that the curve is {\em generically adjoined} to the singularity.

\FIGURE[h]{
\centering
\includegraphics[width=12cm]{stringtwist.pstex}
\put(-37,10){\tiny $Q_{m,2}$}
\put(0,35){\tiny $Q_{m+1,2}$} 
\caption{The CMS becomes twisted as we rotate the radial section. This allows some states to wrap around before decaying. In the shaded circles the CMS has
fixed but undetermined topology.}
\label{stringtwist}}

Fig.\ \hspace{-3pt}(\ref{stringtwist}) shows the effect of generic adjoinment
on a CMS. Whatever the curve does locally around each singularity pair, if
it touches one singularity from each arm 
then globally it will become twisted. There will
exist, superposed, a curve with more twists and with the same type of decay but with all three states raised to a higher position on their tower 
(actually for one twist $n\mapsto n+2$).
If we define states 
\[Q_{m,1}^{n}=Q_{3-}^{n}N_{1}^{m},\hspace{15pt} Q_{m,2}^{n}=Q_{1}^{n}N_{2}^{m}, \]
then one can also see from the diagram that states of high $|m|$ will exist 
close to the singularity pairs.  
The charges of these states and analogous ones are:
\begin{eqnarray}
Q_{2m,1}^{n}   & = & (\a{3},-(2m+1)\a{1}+n\a{3}), \n\\
Q_{2m-1,1}^{n} & = & (\a{2},2m\a{1}+n\a{2}), \n\\
Q_{2m,2}^{n}   & = & (\a{1},(n-1)\a{1}+2m\a{3}), \n\\
Q_{2m-1,2}^{n} & = & (\a{3},-\a{1}+n\a{3}-2m\a{2}), \n\\
Q_{2m,3-}^{n}\!\!\!   & = & (\a{2},-(2m)\a{3}+n\a{2}), \n\\
Q_{2m-1,3-}^{n}\!\!\! & = & (\a{1},-2m\a{3}+(n-1)\a{1}). 
\end{eqnarray}

We have overdefined the states by a factor of two. The $Q_{m,i}$ need only even
$m$ to describe them. The odd ones are equivalent under the relation
\[ Q_{2m,i}^{n}=Q_{-(2m+1),i+1}^{n-2m}\,\,\, , \hspace{20pt}i\in \{ 1,2,3-\} .\]
Note that $Q_{3-}^{n}=Q_{0,1}^{n}=Q_{-1,2}^{n}$ and $Q_{3+}^{n}=Q_{-2,1}^{n}
=Q_{1,2}^{n+2}$.

These $Q_{2m,i}^{n}$ have the same $(g,q)$-charges as the multiplets of higher spin (specifically, as spin multiplets with maximal spin $\frac{3}{2}$, 
$\frac{5}{2}$, $\frac{7}{2}$, $\dots$) as defined in \cite{jerome2} which we know to exist at weak coupling. Of course we doubt that they also possess similar higher
Lorentz quantum numbers --- {\it i.e.}\ that passing through a
 branch cut can notch up or down a states' spin --- but other strange things
can happen. The analysis depends only on the $(g,q)$ charge of the state, so 
even (as we believe) if they are not the same, these states will behave identically for our purposes. 

The other half of the general spin states present at the weak-coupling 
boundary, those with integer spin, we 
call $R$'s.
\begin{eqnarray}
R_{2m,1}^{n}   & = & (\a{3},-(2m)\a{1}+n\a{3}), \n\\
R_{2m-1,1}^{n} & = & (\a{2},(2m-1)\a{1}+n\a{2}), \n\\
R_{2m,2}^{n}   & = & (\a{1},(n-1)\a{1}+(2m-1)\a{2}), \n\\
R_{2m-1,2}^{n} & = & (\a{3},-\a{1}+n\a{3}-(2m-1)\a{2}), \n\\
R_{2m,3-}^{n}\!\!\!   & = & (\a{2},-(2m-1)\a{3}+n\a{2}), \n\\
R_{2m-1,3-}^{n}\!\!\! & = & (\a{1},-(2m-1)\a{3}+(n-1)\a{1}). 
\end{eqnarray}
These transform amongst themselves under the classical monodromy group (let alone under the subgroup of this we get when we include CMS's). They do not include any 
states which become massless, limiting our ability to describe the CMS's upon which they decay.

We now have the labels to state succinctly the BPS spectrum at both weak coupling
points: 
\[
{\rm At}\hspace{5pt}
\left\{ \begin{array}{c}u=\infty \\ u=0 \end{array}\right\}
{\rm we\hspace{5pt} have}\hspace{5pt} Q_{2m,i}^{n}\hspace{5pt} \forall i\in\{ 1,2,3- \} ,
m, n\in {\dbl Z}, 
\left\{ \begin{array}{c}m\geq0 \\ m < 0 \end{array}\right\} .
\] 
\[
{\rm At}\hspace{5pt}
\left\{ \begin{array}{c}u=\infty \\ u=0 \end{array}\right\}
{\rm we\hspace{5pt} have}\hspace{5pt} R_{2m,i}^{n}\hspace{5pt} \forall i\in\{ 1,2,3- \} ,
m, n\in {\dbl Z}, 
\left\{ \begin{array}{c}m > 0 \\ m < 0 \end{array}\right\} .
\]

There are a number of possible types of decay curves, ones involving three 
$Q$'s ($G$'s (and also $L$'s)), involving $QQR$ (the $F$'s), $QRR$ (the $E$'s)
and $RRR$ (the $D$'s).

In regions of weak coupling the higher spin multiplets have larger charges and
are hence more massive. We would therefore expect them to decay into smaller, lighter multiplets. Indeed, the sum of the (maximal) spins of decay product states 
would be equal to the (maximal) spin of the original one. This implies neither of the product states are of higher spin than the original. The decay of a multiplet with highest spin $m$ can thus happen in a variety of ways according to the various partitions of $m$.
The way we believe it to happen is that the spin decreases one-half-at-a-time on a series of curves $F$ where the heavy multiplet necessarily alternates between having integer or half-integer maximal spin as it cascades down, $Q \rightarrow R \rightarrow Q \rightarrow R \dots$, while giving off a $Q_{0,i}$ each time, and terminating also in a $Q_{0,i}$. We also allow for this queue to be jumped by $R$'s decaying into two lower $R$'s, the sum of whose spins equal the original. We do not believe the $Q$'s jump the queue into lower $Q$'s or lower $R$'s, but it would make little difference if it turned out to be the case.      

\FIGURE[h]{
\centering
\includegraphics[width=10cm]{salve1.pstex}
\put(-52,88){$Q_{2m,1}^{n}$}
\put(-39,57){$Q_{2m,2}^{n-2}$}
\put(-2,148){$Q_{2m,3}^{n-4}$}
\put(-67,147){$Q_{2m,1}^{n-6}$}
\put(-270,125){$Q_{-2(m+1),1}^{n-2(m+1)}$}
\put(-230,98){$Q_{-2m,2}^{n-2m}$}
\put(-280,143){$Q_{-2m,3}^{n-2m-6}$}
\put(-282,170){$N_{1}$}
\put(-282,90){$N_{2}$}
\caption{A typical curve $F$.}
\label{salve1}}

The uncomplicated shape of a typical curve $F$ is shown in fig.\ \hspace{-3pt}(\ref{salve1}) --- it justs wraps infinitely many times around the tori on which the trefoils lie. Each $F$ is associated with a particular spin. Although 
the labels change slightly while the curve is in-between the classical branch 
cuts, the spin remains the same. On the outer region, passing either a 
$Q_{2m,i}^{n}$ or an $R_{2m,i}^{n}$ up through both classical branch cuts takes
$i\mapsto i+1$ and $n\mapsto n-2$. After three cycles one returns to the same tower $(2m,i)$ but $n\mapsto n-6$. A similar thing happens if one starts from the inner region, but the notation is not so transparent. Now the towers labelled by subscripts $(2m-2,1)$ have the same spin as $(2m,2)$ and $(2m,3)$ (as states of the same spin are now related by $U_{3+}$ instead of $U_{3-}$) and these three towers rotate into each other as expected. We therefore need only consider the
decay for one label $i$ as the cases for the other two $i$ will be generated by following the curve around one cycle in either direction (of course the decay products also have their ${\dbl Z}_{3}$ label $i$ incremented or decremented). 
To cover all $n$ in 
each tower, we shall need six copies of each curve, one for each $n \bmod 6$.    
We shall also need two types of curve, one for a $Q$ to an $R$ with one half less maximal spin (plus a low $Q$), and one for an $R$ to a $Q$ with one half less 
spin (plus a low $Q$).
There turn out to be two equally valid curves of each type:
\begin{eqnarray*}
Q_{2m,1}^{n} & \longrightarrow & R_{2m,2}^{n-2m} + Q_{0,3}^{n-2m+1} \\
Q_{2m,1}^{n} & \longrightarrow & R_{2m,3}^{n+2m-1} + Q_{0,2}^{n-1} 
\end{eqnarray*}
and
\begin{eqnarray*}
R_{2m,1}^{n} & \longrightarrow & Q_{2(m-1),2}^{n-2m+1} + Q_{0,3}^{n-2(m-1)} \\
R_{2m,1}^{n} & \longrightarrow & Q_{2(m-1),3}^{n+2(m-1)} + Q_{0,2}^{n-1} 
\end{eqnarray*}

The $D$ curves which allow multiplets with integer maximal spin $m$ to `jump the queue' also have the same simple shape as the $F$'s. There is one for each bipartition of $m$, namely for some $m=k+l$ 
\[ R_{2m,1}^{n} \longrightarrow R_{2k,2}^{n-2k} + R_{2l,3}^{n+4k}. \]
We may now forget about the higher spin multiplets and the $D$'s and $F$'s and return our focus to the CMS's at stronger coupling.

\subsection{The Global Curves}

We are now in a position to inspect how the classical CMS splits as we include
non-perturbative quantum corrections. An infinite number of decays occur on the
classical CMS, ones for all $Q_{3\pm}^{n}$. We find the curve does not just 
bifurcate, but that each piece also `unravels'.
We begin by looking at two test segments, and find that under twisting they 
generate the whole of the $G$'s.

It could be said that the role of the classical CMS was to partition each of 
the sectors into two domains. This must remain the job of the $G$'s, which 
coalesce to form the single CMS as we turn off the non-perturbative effects.
With this in mind, let us begin with two curves whose presence is staring us 
in the face, namely the decay of $Q_{3-}^{1}$ into $Q_{1}^{1}+Q_{2}^{1}$, which
will touch the top singularity of each arm in the 1,2 sector, and 
$Q_{3-}^{2}\rightarrow Q_{1}^{2}+Q_{2}^{2}$ which will touch the bottom of 
each pair. These are displayed in fig.\ \hspace{-3pt}(\ref{primary}).  
\FIGURE[ht]{
\centering
\includegraphics[width=13cm]{g00.pstex}
\put(-260,73){\scriptsize $Q_{1}^{2}$}
\put(-274,86){\scriptsize $Q_{1}^{0}$}
\put(-274,100){\scriptsize $Q_{1}^{1}$}
\put(-260,2){\scriptsize $Q_{2}^{2}$}
\put(-274,15){\scriptsize $Q_{2}^{0}$}
\put(-274,29){\scriptsize $Q_{2}^{1}$}
\put(-49,72){\scriptsize $Q_{1}^{2}$}
\put(-56,86){\scriptsize $Q_{1}^{0}$}
\put(-56,100){\scriptsize $Q_{1}^{1}$}
\put(-42,1){\scriptsize $Q_{2}^{2}$}
\put(-56,15){\scriptsize $Q_{2}^{0}$}
\put(-56,29){\scriptsize $Q_{2}^{1}$}
\put(2,46){\small $Q_{3-}^{2}$}
\put(-110,46){\small $Q_{1}^{2} + Q_{2}^{2}$}
\put(-212,50){\small $Q_{3-}^{1}$}
\put(-327,50){\small $Q_{1}^{1} + Q_{2}^{1}$}
\put(-140,64){\small $N_{1}$}
\put(-140,-8){\small $N_{2}$}
\put(-357,64){\small $N_{1}$}
\put(-357,-8){\small $N_{2}$}
\put(-26,77){\large $G_{3-}^{2}$}
\put(-232,82){\large $G_{3-}^{1}$}
\caption{CMS's which will generate all the $G$'s. The states becoming massless
at each singularity are labelled.}
\label{primary}}

As we rotate the 2-dimensional section the curve 
twists. $U$-covariance demands that if after a half twist in the positive sense (such that the other arm is now on top and we are in the 3,1 section), we superimpose this curve back on the original section whilst multiplying by $UU_{3-}$ ({\it i.e.}\ cyclically permuting back the indices), then this curve must also exist. 
We find it to be 
fig.\ \hspace{-3pt}(\ref{secondary}).

\FIGURE[h]{
\centering
\includegraphics[width=13cm]{g01.pstex}
\put(-260,73){\scriptsize $Q_{1}^{2}$}
\put(-260,2){\scriptsize $Q_{2}^{2}$}
\put(-274,15){\scriptsize $Q_{2}^{0}$}
\put(-260,29){\scriptsize $Q_{2}^{1}$}
\put(-47,72){\scriptsize $Q_{1}^{2}$}
\put(-56,86){\scriptsize $Q_{1}^{0}$}
\put(-56,100){\scriptsize $Q_{1}^{1}$}
\put(-42,1){\scriptsize $Q_{2}^{2}$}
\put(-56,18){\scriptsize $Q_{2}^{0}$}
\put(-56,29){\scriptsize $Q_{2}^{1}$}
\put(-80,46){\small $Q_{3+}^{2}$}
\put(-28,46){\small $Q_{1}^{4} + Q_{2}^{2}$}
\put(-332,50){\small $Q_{3+}^{1}$}
\put(-265,50){\small $Q_{1}^{3} + Q_{2}^{1}$}
\put(-306,86){\scriptsize $Q_{2}^{1}$}
\put(-275,86){\scriptsize $Q_{3}^{1}\!-Q_{1}^{1}$}
\put(-140,64){\small $N_{1}$}
\put(-140,-8){\small $N_{2}$}
\put(-357,64){\small $N_{1}$}
\put(-357,-8){\small $N_{2}$}
\put(-50,57){\large $G_{3+}^{2}$}
\put(-284,64){\large $G_{3+}^{1}$}
\caption{Rotation of fig.\ \hspace{-3pt}(\ref{primary}).}
\label{secondary}}

\FIGURE[ht]{
\centering
\includegraphics[width=14cm,height=6cm]{g02.pstex}
\put(-357,81){\tiny $Q_{3-}^{-1}$}
\put(-390,81){\tiny $Q_{1}^{-1}Q_{2}^{-1}$}
\put(-380,101){\tiny $Q_{3+}^{-1}$}
\put(-362,101){\tiny $Q_{1}^{1}Q_{2}^{-1}$}
\put(-369,29){\tiny $Q_{-2,2}^{-1}$}
\put(-348,29){\tiny $Q_{2}^{1}Q_{3}^{-1}$}
\put(-347,5){\tiny $Q_{1}^{1}$}
\put(-374,5){\tiny $Q_{3}^{1}Q_{2}^{1}$}
\put(-388,125){\tiny $Q_{2}^{-1}$}
\put(-369,125){\tiny $Q_{1}^{1}Q_{3+}^{-1}$}
\put(-279,77){\tiny $Q_{3-}^{0}$}
\put(-307,77){\tiny $Q_{1}^{0}Q_{2}^{0}$}
\put(-300,104){\tiny $Q_{3+}^{0}$}
\put(-282,104){\tiny $Q_{1}^{2}Q_{2}^{0}$}
\put(-308,29){\tiny $Q_{1}^{2}$}
\put(-292,29){\tiny $Q_{3}^{2}Q_{2}^{2}$}
\put(-176,83){\tiny $Q_{3-}^{1}$}
\put(-204,83){\tiny $Q_{1}^{1}Q_{2}^{1}$}
\put(-228,103){\tiny $Q_{3+}^{1}$}
\put(-208,103){\tiny $Q_{1}^{3}Q_{2}^{1}$}
\put(-223,136){\tiny $Q_{2}^{1}$}
\put(-205,136){\tiny $Q_{3}^{1}Q_{1}^{1}$}
\put(-90,82){\tiny $Q_{3-}^{2}$}
\put(-120,82){\tiny $Q_{1}^{2}Q_{2}^{2}$}
\put(-137,103){\tiny $Q_{3+}^{2}$}
\put(-117,103){\tiny $Q_{1}^{4}Q_{2}^{2}$}
\put(-142,149){\tiny $Q_{2}^{2}$}
\put(-126,149){\tiny $Q_{3}^{2}Q_{1}^{2}$}
\put(-10,83){\tiny $Q_{3-}^{3}$}
\put(-39,83){\tiny $Q_{1}^{3}Q_{2}^{3}$}
\put(-56,102){\tiny $Q_{3+}^{3}$}
\put(-36,102){\tiny $Q_{1}^{5}Q_{2}^{3}$}
\put(-61,154){\tiny $Q_{2}^{3}$}
\put(-43,154){\tiny $Q_{3-}^{3}\!\!Q_{1}^{3}$}
\put(-37,142){\tiny $Q_{1,1}^{3}$}
\put(-63,142){\tiny $Q_{1}^{1}Q_{3}^{3}$}
\put(-45,133){\tiny $Q_{3-}^{3}$}
\put(-32,133){\tiny $Q_{2}^{3}Q_{1,1}^{3}$}
\put(-35,38){\tiny $Q_{1,1}^{3}$}
\put(-64,38){\tiny $Q_{3+}^{3}\!Q_{2}^{1}$}
\put(-53,58){\tiny $Q_{1}^{3}$}
\put(-41,46){\tiny $Q_{3+}^{1}\!Q_{2}^{1}$}
\caption{More $G_{3-}$ and $G_{3+}$'s. The pattern continues to the left and right.}
\label{g02}}

Repeated twists of the arms around each other in this manner suffices to 
generate from just the two curve segments in fig.\ \hspace{-3pt}(\ref{primary}) all the curves (or rather all parts of the same two curves) necessary to 
partition the region into two domains each carrying one of the towers $3-$ or 
$3+$. A few
of these are shown in fig.\ \hspace{-3pt}(\ref{g02}).

\subsection{The Cores}

In the 2D-sections we have drawn, each arm looks remarkably like a copy of the 
SU(2) moduli space. It has two singularities, a quantum branch cut between them, 
and a classical one tending off to infinity. SU(3) has three natural SU(2) embeddings, one for each root. As each of our arms are also associated with one of the 
three roots, it is natural to identify the region near the two singularities of
 an arm as the place where the SU(3) coupling is strong only in the projection
onto the relevant SU(2) subspace.
To complete this picture we also need the CMS corresponding to the unique decay 
curve in the pure SU(2) case. For the arm labelled by $i$, which is associated to 
the 
$\a{i}$ direction and has classical and quantum monodromies $N_{i}$ and 
$M_{2i}$ respectively, we find a CMS which is topologically a circle 
passing through
both of the singularity pair. On the left of the quantum branch cut all the 
states $Q_{i}^{n}$ decay to a combination of $Q_{i}^{0}$ and $Q_{i}^{1}$. On 
the right of the cut they decay to $Q_{i}^{1}$ and $Q_{i}^{2}$.
This is shown in fig.\ \hspace{-3pt}(\ref{sem10}).
We call this CMS $L_{i}$. Returning to the whole picture of the trefoil pair, 
we see that the $L_{i}$ form one continuous closed tube $L$ in the shape of a 
trefoil, lined on the top and bottom by the two trefoils that are the two 
singularities, as seen in fig.\ \hspace{-3pt}(\ref{lissy}).

\FIGURE[h]{
\centering
\makebox[12cm]{\includegraphics[width=8cm]{sem10.pstex} 
\put(-23,110){$Q_{i}^{n}$, $W_{i}$}
\put(-167,109){$Q_{i}^{n}$, $W_{i}$}
\put(-42,66){$Q_{i}^{1}$}
\put(-90,66){$Q_{i}^{1}$}
\put(-42,33){$Q_{i}^{2}$}
\put(-90,33){$Q_{i}^{0}$}
\put(-183,20){$N_{i}$}
\put(-65,96){\scriptsize $Q_{i}^{1}$}
\put(-60,0){\scriptsize $Q_{i}^{2}$}
\put(-73,13){\scriptsize $Q_{i}^{0}$}}
\caption{The region where the $\a{i}$ direction is strongly coupled.} 
\label{sem10}}

\FIGURE[h]{
\centering
\makebox[12cm]{\includegraphics[width=12cm,height=6cm]{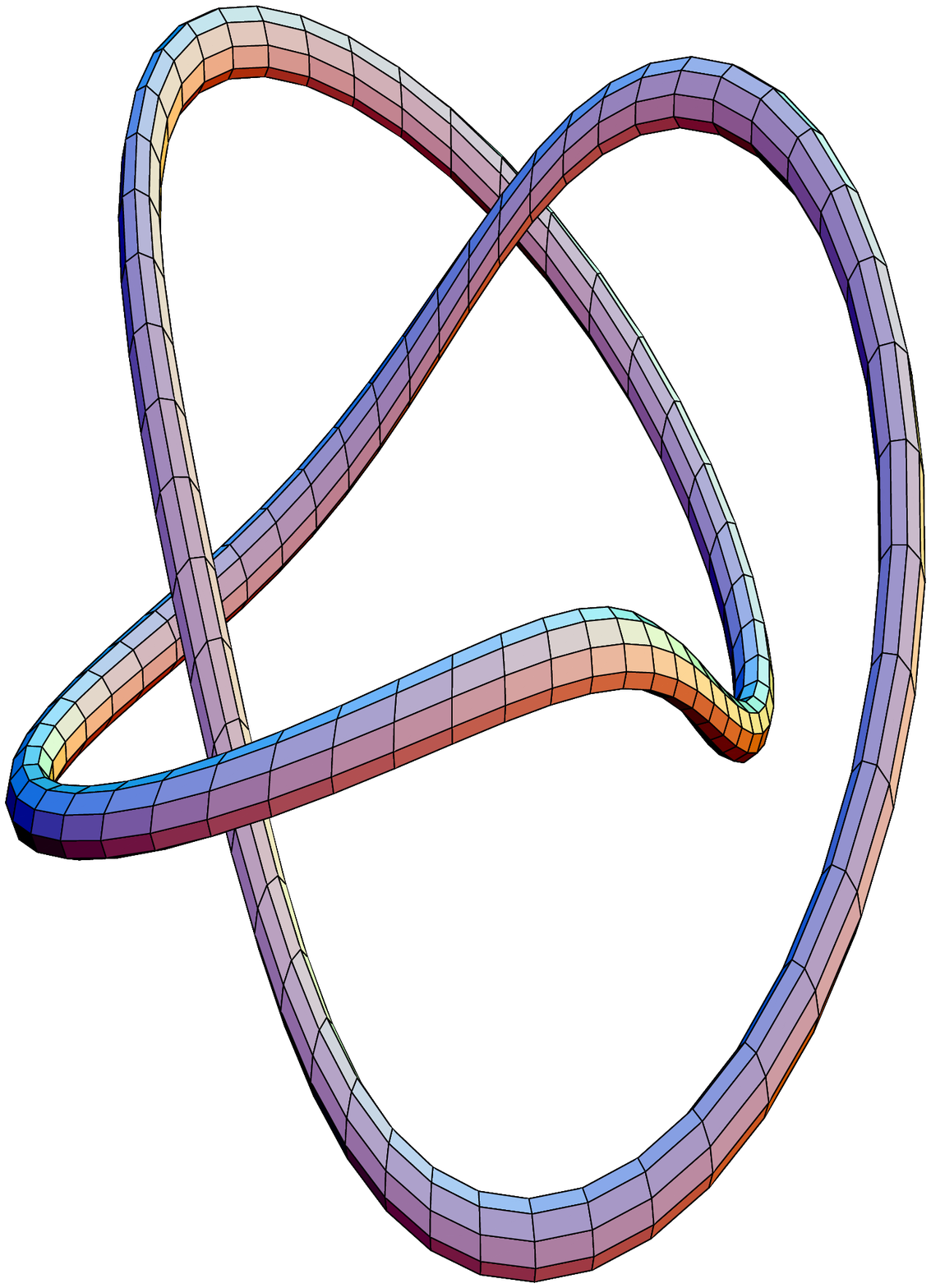}}
\caption{The CMS $L$ which bounds the core.}
\label{lissy}}

If we concentrate on the $\a{1}$ arm, we notice that the $Q_{2}$ and $Q_{3}$ towers 
are not blocked from $M_{2}$ by $L_{1}$, nor are the $Q_{m,1}$ which are these states after being wrapped around the $\a{1}$ singularity pair a number of 
times. Luckily the $G$'s fulfill an unexpected dual role, limiting the states
getting inside $L$ to exactly those which transform nicely through the quantum branch cut there.
From now on we shall refer to this region inside $L$ as the {\it core}.  

After the decay of the higher spin multiplets we are left with three towers at each point,
$Q_{1}$, $Q_{2}$, and $Q_{3+}$ or $Q_{3-}$. 
If we are considering the core associated to the $\a{1}$ arm, we think of these
towers as $Q_{1}$, $Q_{-1,1}$, $Q_{-2,1}$ and $Q_{0,1}$; and for that with the $\a{2}$ arm,
 $Q_{0,2}$, $Q_{2}$, $Q_{1,2}$ and $Q_{-1,2}$ respectively.
When a curve twists a lot of times around a core, one finds another cascade of decays. Here what happens is that the $n$ index, which distinguishes states within a
tower, is preserved, but $m$ increases 
(or decreases if the curve wraps in the other sense) until
it has reached the lightest state it can with that value of $n$. 
This occurs for
\[ n = 2m+1, \hspace{10pt} n=2m+2, \hspace{10pt} n=2m+3 \] 
We label these states $S_{m,i}^{-1}$,  $S_{m,i}^{0}$ and $S_{m,i}^{1}$ 
respectively.
For all the intermediate states in the cascade, we find they are confined to a strip bounded both away from the core and from the weak coupling regions, and 
by the classical branch cut in a way shown in figs.\ \hspace{-3pt}(\ref{k},\ref{l}).
The other decay products are all $Q_{1}$'s which we know to be excluded from the core by $L_{1}$.
The special states $S_{m,i}$ are confined merely to strips which end at the 
core.
$S_{m,i}^{0}$ is invariant through the quantum monodromy in the core, and the
other two sets map into each other by the action of the monodromy, the 
$S_{m,i}^{-1}$ 
on the left and $S_{m,i}^{1}$ on the right. 

\FIGURE[h]{
\centering
\includegraphics[width=13cm]{betterroll.pstex}
\put(-378,91){$Q_{m+2,i}^{2(m+2)+2}$}
\put(-248,91){$Q_{m+1,i}^{2(m+1)+4}$}
\put(-60,87){$Q_{m,i}^{2m+6}$}
\caption{A sketch of the existence domains of some states around a core.}
\label{k}}
\FIGURE[h]{
\centering
\makebox[8cm]{\includegraphics[width=6cm]{broll2.pstex}
\put(-200,117){$Q_{m+1,i}^{n}$}
\put(-145,117){$Q_{m,i}^{n}$}
\put(-110,36){$Q_{m+1,i}^{n}$}
\put(-154,36){$Q_{m,i}^{n}$}
\put(-82,117){$Q_{m,i}^{n}$}
\put(-34,117){$Q_{m-1,i}^{n}$}
\put(2,36){$Q_{m,i}^{n}$}
\put(-5,17){$+Q_{i}^{4m+2-n}$}
\put(-48,35){$Q_{m-1,i}^{n}$}}
\caption{The decays involved in an existence domain strip.}
\label{l}}

We now see that we have 3 towers of states within each core $L_{i}$ 
(though only two at any point), the $S_{m,i}^{-1}$, $S_{m,i}^{0}$ and $S_{m,i}^{1}$. This time the label is on the bottom, and we see that in a sense $n$ and $m$ have swapped roles. These are strong coupling towers, where the separation of charge is not that of a gauge boson, but of the difference between two of 
these. For each $i$, one might expect a BPS state to exist with such charge, which we believe to be the case, and later we explain their necessity within the model.
We expect that such a tower can be seen to be formed from a dual model, weakly 
coupled at this core, with the charge angle in the monopole 
moduli space translating back to lie in a different direction within SU(3).

\subsection{The Meeting of the Cores}

Even for the classical case, the trefoil degenerates at a certain $S^{3}$ 
radius $R=\sqrt{3}\Lambda$, its innermost parts pinching together while the knot `unties', and 
after which three unlinked circles remain.
As we decrease the $S^{3}$ radius in the full quantum theory, the distance 
between the two trefoils increases, as does the core between them. At a 
critical radius something gives --- the trefoils pass through each other at three
points. Also, at this same radius, both trefoils degenerate at their centre.
Further decreasing the radius, the intersection with the singular surface
takes the form of six circles, two for each root, still with a core between 
them.  

One can check that the top singularity of the pair associated to the bottom
arm may pass with impunity through the quantum cut of the top arm, between its singularity pair. This is what happens for $R$ only a little below 
the critial value. The cores in this case must wrap and tangle, or merge, as seen in fig.\ \hspace{-3pt}(\ref{choose}).
That they merge is by far the most elegant idea. We have been unable 
to find consistent models incorporating wrap and tangle and believe them to 
have generic problems.
We thus consider the picture presented in the bottom half of 
fig.\ \hspace{-3pt}(\ref{choose}). One can see that the set of $G$ curves 
wrapped around each core are necessarily squeezed to a point, then 
bisected. We call a point through which an infinite number of CMS's pass
an {\it accumulation point}.

\FIGURE[h]{
\centering
\includegraphics[width=9cm]{tubemerge2.pstex}
\includegraphics[width=13cm]{tubemerge3.pstex}
\caption{Two possible scenarios for when two roots become 
strongly coupled.}
\label{choose}}

\subsection{Accumulation Points}

We remarked before that a CMS depends on the lattice generated by integer
linear combinations of any two of the three states involved in a decay across it. Conversely, any two charges give a potentially relevant CMS. 
We label the equivalence classes of CMS's, or two-generator lattices as 
$[(g_{1},q_{1}),(g_{2},q_{2})]$.

In this notation $L_{i}= [ (0,\a{i}),(\a{i},0)] $, for instance.
Other curves of subsequent importance are
\[ \beta=[(0,\a{i}),(0,\a{j})] , \hspace{10pt} 
\gamma_{ij}(n,n^{\prime}) = [ Q_{i}^{n},Q_{j}^{n^{\prime}} ]. \]
We label
\[ \gamma_{ij}=\bigcap_{n,n^{\prime}}{\gamma_{ij}(n,n^{\prime})}. \]
Although we can say little about where two curves cross, if they do cross then
the combined constraints may imply that other curves also pass through this 
point too.

From fig.\ \hspace{-3pt}(\ref{choose}) we can see that there must be points 
where $L_{1}$ crosses $L_{2}$. It would be helpful if all the wrapped $G$'s 
could then be shown to accumulate here too. $L_{1}\cap L_{2}$ is not 
sufficient, but $L_{1}\cap L_{2}\cap\beta$ is equivalent to 
$L_{1}\cap L_{2}\cap\gamma_{12}$ which is very strong indeed.

If the cores merge then it might seem that $L_{1}$ and $L_{2}$ have a boundary 
on the accumulation points where they meet. This cannot be the case, so we
must assume that we have so far overlooked components of these curves 
encircling the `wrong' cores. 
All these curves act within a tower, mapping elements of the tower into the
two with the lowest value of $n$. Towers are labelled by subscripts $m$ and 
$i$.
As $L_{1}$ encircles the $\a{1}$ core it falls
back on itself to form a closed curve as in the SU(2) case. 
Wrapping once, we find the relation $L_{3}=L_{2}N_{1}$,
and continued wrapping will create a different curve with each rotation. Those 
with an even number of cycles from $L_{3}$ behave differently to those with
an even number of cycles from $L_{2}$. In the former case
\[ \forall n,\,\,\, Q^{n-2m}_{(2m+1),1}\longrightarrow Q_{-(2m+1),1}^{1}
+Q_{-(2m+1),1}^{0} \Leftrightarrow \forall n, \,\,\, 
[W_{2},Q^{n-2m}_{-(2m+1),1}] \]
If, as is likely, any two of these meet, then they all meet, at $[\beta\cap 
L_{2}]$. Similarly for an odd number of cycles (those associated with the 
tower $Q_{2m+1,1}$), they all meet at  $[\beta\cap 
L_{3}]$. Odds cross evens at different points in all cases.
Finally, just as $L_{2}$ is a special case of one of the family wrapping around
the $\a{1}$ core, associated to the $Q_{2m,1}$'s, then $L_{1}$ is a special 
case of a family around the $\a{2}$
core, associated to the $Q_{2m,2}$'s, which must also be boundaryless when 
the cores merge. Therefore we must 
consider $L_{1}$ to be a special case of this family, which acts slightly differently around the $\a{1}$ core. Here we find that a rotating around once maps 
$(2m,2)\mapsto 
(-2m,2)$, and so performing two cycles maps one back to the same tower.
We might still expect an accumulation point where $L_{1}$ (associated to 
$Q_{0,2}$) would intersect the curves associated to all the other $Q_{2m,2}$'s,
which would happen when $\beta\cap L_{1}$.
Maybe this does not occur until the cores merge, however.
The addition of all these curves also shields the core $L_{i}$ from the other 
two gauge bosons $W_{j}$ and $W_{k}$.

\FIGURE[ht]{
\centering
\includegraphics[width=6cm]{newwraps.pstex}
\put(-170,30){$N_{1}$}
\put(-70,150){$(2m+1,1)$}
\put(-18,135){$(2m,1)$}
\put(-97,90){$(2m,2)$}
\put(-52,115){$L_{1} \cap \beta$}
\put(-160,84){$L_{2} \cap \beta$}
\put(-85,26){$L_{3} \cap \beta$}
\caption{The curves encircling the $\a{1}$ core.}
\label{newwraps}}
     
When the cores first touch, we envisage that all six accumulation points
coalesce as in fig.\ \hspace{-3pt}(\ref{2core1}). Thus we have 
$(\beta\cap L_{1})\cap(\beta\cap L_{2})\cap(\beta\cap L_{3})$. This then 
guarantees that all the wrapped $G$'s (and more besides!) will accumulate at
this point. After this, the six points separate into two sets of three, each
with the same conditions. While the cores pass through each other, these 
accumulation points rotate around the merged core until, at the point where the
cores separate again, the two points have swapped position, top-for-bottom and
bottom-for-top.

\FIGURE[ht]{
\centering
\makebox{\includegraphics[width=13cm]{2core1.pstex}
\put(-355,70){\tiny 2}
\put(-351,78){\tiny 2}
\put(-347,70){\tiny 3}
\put(-343,78){\tiny 3}
\put(-335,70){\tiny 1}
\put(-331,78){\tiny 1}
\put(-253,68){\tiny 1}
\put(-248,60){\tiny 1}
\put(-246,68){\tiny 3}
\put(-240,60){\tiny 3}
\put(-262,40){\tiny 2}
\put(-257,44){\tiny 2}
\put(-137,70){\tiny 2}
\put(-134,78){\tiny 2}
\put(-126,70){\tiny 3}
\put(-121,78){\tiny 3}
\put(-109,70){\tiny 1}
\put(-105,78){\tiny 1}
\put(-32,68){\tiny 1}
\put(-23,60){\tiny 1}
\put(-10,68){\tiny 3}
\put(-4,60){\tiny 3}
\put(-41,40){\tiny 2}
\put(-37,44){\tiny 2}
\put(-135,50){\tiny 3}
\put(-129,44){\tiny 3}
\put(-121,50){\tiny 2}
\put(-111,44){\tiny 2}
\put(-75,35){\tiny 1}
\put(-69,35){\tiny 3}}
\caption{The accumulation points merge as the cores touch.}
\label{2core1}}
  
\subsection{Double Cores}

The merging of cores presents us with new problems to resolve. Clearly, without
the addition of more curves, we find ourselves with the union, not the 
intersection of the spectra within the cores $L_{i}$ and $L_{j}$. More 
worryingly, the strong coupling towers of states $S_{m,i}$ and $S_{m,j}$ do
not transform well through the opposite quantum monodromies, as seen in fig.\ 
\hspace{-3pt}(\ref{b1}). 

\FIGURE[h]{
\centering
\includegraphics[width=12cm]{branle1.pstex}
\put(-280,140){\scriptsize $Q_{1}^{0}$}
\put(-252,140){\scriptsize $Q_{1}^{2}$}
\put(-280,123){\scriptsize $Q_{1}^{1}$}
\put(-252,123){\scriptsize $Q_{1}^{1}$}
\put(-296,105){\scriptsize $S_{2m+1,1}^{-1}$}
\put(-252,105){\scriptsize $S_{2m,1}^{1}$}
\put(-296,88){\scriptsize $S_{2m,1}^{0}$}
\put(-252,88){\scriptsize $S_{2m,1}^{0}$}
\put(-296,70){\scriptsize $S_{2m,1}^{-1}$}
\put(-252,70){\scriptsize $S_{2m+1,1}^{1}$}
\put(-296,53){\scriptsize $S_{2m+1,1}^{0}$}
\put(-252,53){\scriptsize $S_{2m+1,1}^{0}$}
\put(-313,111){?}
\put(-82,105){\scriptsize $Q_{2}^{-1}$}
\put(-54,105){\scriptsize $Q_{2}^{1}$}
\put(-82,88){\scriptsize $Q_{2}^{0}$}
\put(-54,88){\scriptsize $Q_{2}^{0}$}
\put(-98,70){\scriptsize $S_{2m+1,2}^{-1}$}
\put(-54,70){\scriptsize $S_{2m,2}^{1}$}
\put(-98,53){\scriptsize $S_{2m,2}^{0}$}
\put(-54,53){\scriptsize $S_{2m,2}^{0}$}
\put(-98,35){\scriptsize $S_{2m,2}^{-1}$}
\put(-54,35){\scriptsize $S_{2m+1,2}^{1}$}
\put(-98,18){\scriptsize $S_{2m+1,2}^{0}$}
\put(-54,18){\scriptsize $S_{2m+1,2}^{0}$}
\put(-22,42){?}
\caption{The states native to each core are alien to the other one.}
\label{b1}}

\FIGURE[t!]{
\centering
\includegraphics[width=12cm]{branle2.pstex}
\put(-240,135){$S_{2m,1}^{1}$}
\put(-246,115){$S_{2m,1}^{0}$}
\put(-259,70){$S_{2m+1,1}^{1}$}
\put(-252,50){$S_{2m+1,1}^{0}$}
\put(-117,119){$S_{2m,2}^{0}$}
\put(-110,100){$S_{2m,2}^{-1}$}
\put(-110,60){$S_{2m+1,2}^{0}$}
\put(-111,41){$S_{2m+1,2}^{-1}$}
\put(-185,112){\scriptsize $(\a{3},\a{3})$}
\put(-183,95){\scriptsize $(\a{3},0)$}
\put(-188,74){$Q_{2}^{1}$}
\put(-186,58){$Q_{2}^{0}$}
\put(-166,68){$Q_{1}^{2}$}
\put(-168,50){$Q_{1}^{1}$}
\put(-215,84){\large $P_{1}$}
\put(-138,80){\large $P_{2}$}
\caption{There is a double core region from which the $S$'s are excluded, but 
four new states are created.}
\label{b2}}

What is more likely is that families of curves form between the accumulation 
points upon which nearly all the states of both cores decay, to leave a region
in the centre with just a few (indeed a finite number) of states which 
transform nicely through both quantum mondromies. We call this central region
a {\it double core}. The new curves are necessarily circular in our 
2D-sections, hence spherical in 3-dimensions and 3-spherical in the full 4 
dimensions. Thus we find three bounded regions for which two root directions 
are strongly coupled, and within each of which lies a double core carrying a
finite number of BPS states.

From fig.\ \hspace{-3pt}(\ref{b2}) one can see that there are eight families of curves required to fulfill this task. In each case one is met with a cascade
of $(m,i)\mapsto (m-2,i)$ giving as a by-product $2P_{i}$. If we define the
massive gauge bosons $W_{1}$, $W_{2}$ and $W_{3}$ as having charges $(0,\a{1})$, 
$(0,\a{2})$ and $(0,-\a{3})$ respectively, then $P_{i}$ has charge 
equal to the difference $W_{i-1}-W_{i+1}$.
Thus the $P_{i}$ seem in some way to perform the role of gauge bosons within
the core. $P_{i}$ is invariant under the monodromies of the $\a{i}$ arm.
Under the other classical monodromies it transforms into the other two $P$'s,
but under the other quantum monodromies it is not so well behaved --- the states 
into which it transforms do not have unit magnetic charge. These rogue states
decay on a set of curves, for each rogue state one curve emanating from the top, 
and 
one from the bottom of the cut across which the rogue state is created.
As all these curves must pass through the accumulation points, they must touch,
and hence bound the existence domain of the rogue state. The decays on the 
curves are back to a $P$ plus the state becoming massless at the singularity
through which the curve passes.

The $P_{i}$ do not exist at weak coupling, and most likely decay to gauge 
bosons on a curve similar to the $D$'s and $F$'s but this time closed after 
three cycles, most probably as 
$P_{i}\rightarrow 2W_{i-1}-W_{i}$. 
Within this curve all three $P_{i}$ exist at every point.
Recall that all the gauge bosons are shielded from the core by the $L_{i}$.

Finally, note that $R_{0,1}^{0}=(\a{3},0)$ and $R_{0,1}^{1}=(\a{3},\a{3})$
are also states which live in the double core but not at weak coupling.
These decay on the curves in fig.\ \hspace{-3pt}(\ref{oo}). 

\FIGURE[ht]{
\centering
\includegraphics[width=13cm]{exitq3q3.pstex}
\put(-300,40){$(\a{3},\a{3})$}
\put(-90,33){$(\a{3},0)$}
\put(-328,63){\scriptsize $Q_{2}^{0}Q_{1}^{1}$}
\put(-120,20){\scriptsize $Q_{2}^{1}Q_{1}^{2}$}
\caption{$R_{0,1}^{0}$ and $R_{0,1}^{1}$ are confined to 3-spheres around the double core.}
\label{oo}}

\subsection{The $A_{2}$ Singularities}

We have dealt with the the double cores which occur around the $A_{1}$
singularities of the singularity surface. There also exist two $A_{2}$ 
singularities at the critical radius $R=\sqrt{3}\Lambda$.
Again the cores will coincide, and again we expect them to merge.
That the coupling is strong in two root directions implies that it must be 
strong in the third root direction too. We do not expect a 
triple core. Therefore we must have three double cores squeezed up against each
 other in the neighbourhood of $R=\sqrt{3}\Lambda$ and $u=0$.
The fact that the right hand side of the $\a{i}$ core always merges with the
left hand side of the $\a{i+1}$ core suggests how the three tubes slide through
 each other merging only pairwise, the detailed exposition of which this 
2-dimensional medium does not lend itself.

\FIGURE[ht]{
\centering
\includegraphics[width=10cm]{5tref2.pstex}
\caption{The $A_{2}$ singularities highlighted from the point of view of the
bottom of the quantum cuts.}}

Further decreasing the radius $R$ now decreases the coupling strength slightly,
although we still retain cores until just after the point when 
$R=\Lambda/\sqrt[6]{2}$ when the intersection with the singularity surface 
becomes empty. While $\Lambda/\sqrt[6]{2} < R < \Lambda\sqrt{3}$ we live in
a region similar in many ways to when $\Lambda\sqrt{3}<R< \infty$, but, if 
anything, with a simpler structure.
The branch cuts are shown in fig.\ \hspace{-3pt}(\ref{lippy}).
Three toroidal cores exist around the quantum cuts.
The $G$'s all pass through the $A_{2}$ singularities, after which they `untie'
globally such that each $G$ is rotationally symmetric about the centre of the 
torus around which it wraps, as shown in fig.\ \hspace{-3pt}(\ref{ion1}).

\FIGURE[h]{
\centering
\includegraphics[width=5cm]{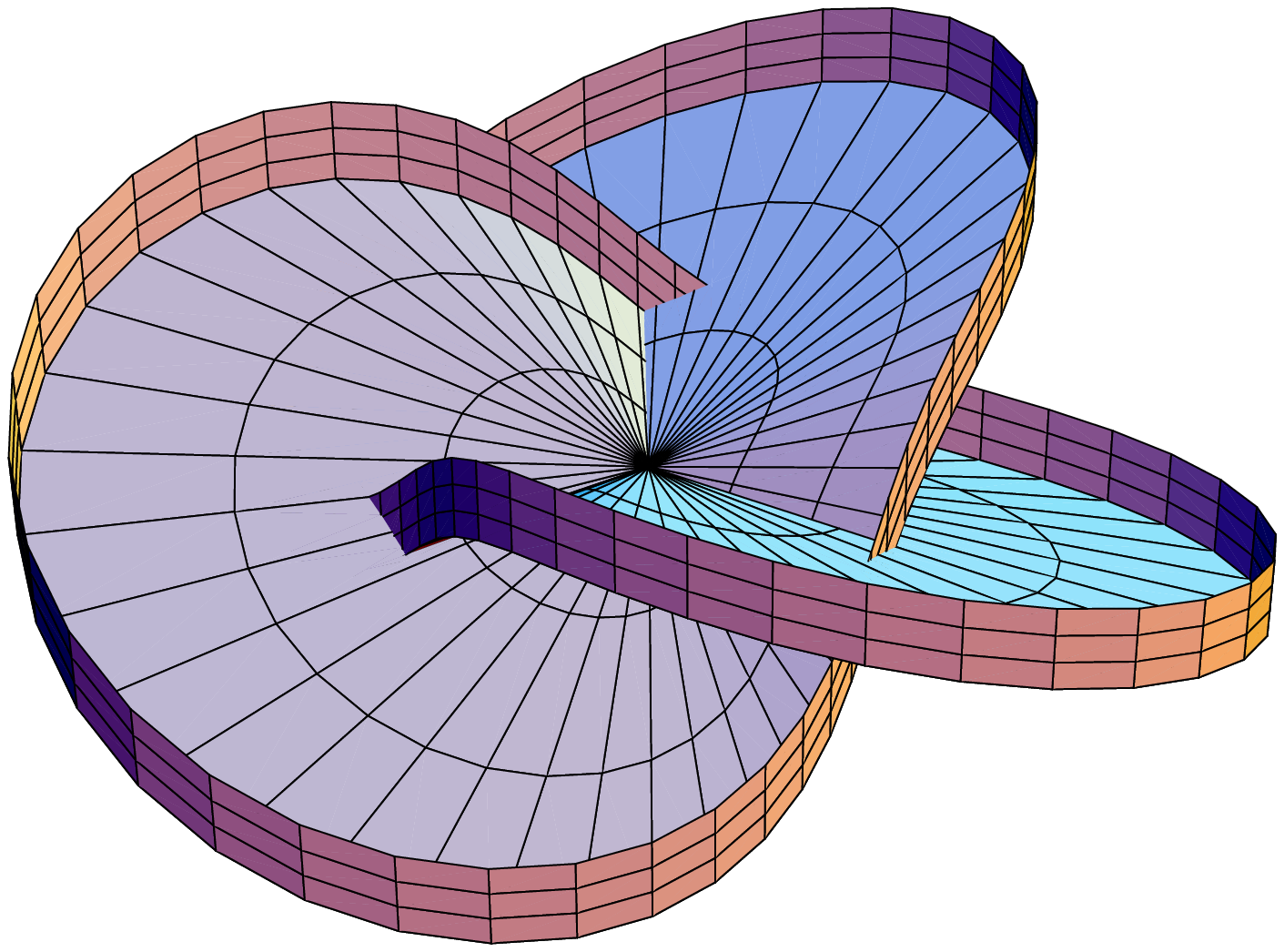}
\includegraphics[width=5cm]{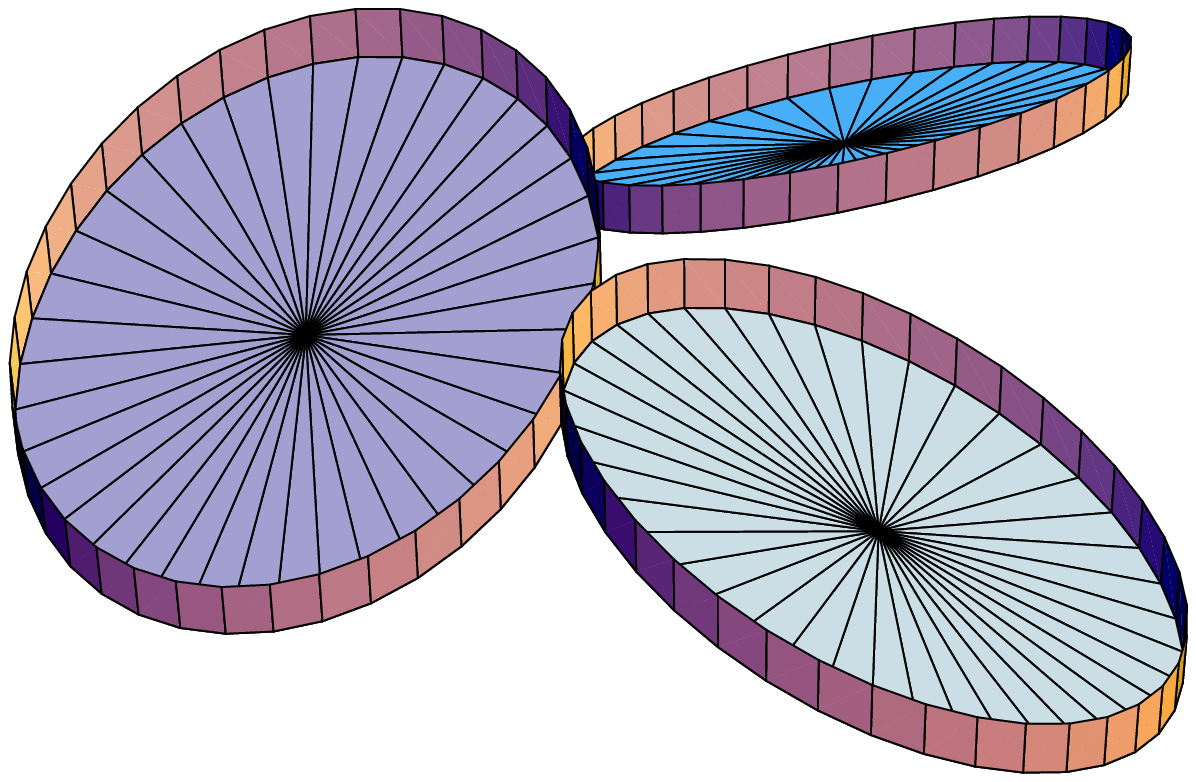}
\caption{The branch cuts for large and for small radii.}
\label{lippy}}

\FIGURE[h]{
\centering
\includegraphics[width=13cm]{ion1.pstex}
\put(-319,67){\scriptsize $Q_{m,1}^{2m+2}$}
\put(-188,67){\scriptsize $Q_{m,1}^{2m+2}$}
\put(-59,67){\scriptsize $Q_{m,1}^{2m+2}$}
\put(-368,28){\scriptsize $N_{1}$}
\put(-120,28){\scriptsize $N_{1}$}
\caption{The local wrapping is mirrored in this sector.}
\label{ion1}}

\subsection{Remarks}

We have found a consistent model of the moduli space of pure SU(3) with no 
matter hypermultiplets, and seen 
that there are disjoint bounded regions called double cores within which
there exist only a finite number of BPS states, all in the form of hyper- and 
vector multiplets. There are none of the three semiclassical gauge bosons in 
these regions, instead there exist states $P_{i}$ 
with charges given by the difference of the charges of two of the gauge bosons.

Our analysis also predicts the existence semiclassically of $Q_{2m,2}^{n}$ and
$Q_{2m,3-}^{n}$ with charges 
$(\a{1},(n-1)\a{1}+2m\a{2})$ and $(\a{2},n\a{2}-2m\a{3})$ respectively. 
We speculate that perhaps these are generated in a similar
manner to the states $Q_{2m,1}^{n}$, whereby they can be seen as the effect
of exciting momentum in the compact $\{ \a{3}-(-\a{2})\} $ and $\{ 
\a{3}-(-\a{1})\} $ 
directions respectively of a higher dimensional monopole moduli space with 
potential, 
a sort of Witten effect which vanishes upon alignment of the two Higgs vevs.
These directions are precisely those of $P_{1}$ and $P_{2}$, and indeed we see 
towers created at strong coupling caused by excitation in these directions
alone, with the charge proportional to the magnetic charge in units 
$\{ 0,\pm 1 \} $ as for SU(2). 

There is an alternative formulation of the BPS states as M2-branes in M-theory
\cite{witm5,henyi,mik,fs} which may shed further light on this problem. Of 
particular interest is how a higher spin multiplet may be interpreted, and the
number of boundaries it might possess, perhaps twice the value of the highest 
spin state in the multiplet.

There also exists an index theorem calculation of Stern and Yi \cite{pyi} which
counts the BPS states for SU($n$). This gives a lower bound on the number of 
states present and could lead to a more definitive statement as to whether or 
not any of these new states occur. In \cite{wolf} Lerche predicts the BPS spectrum at the point $u=v=0$. We cannot reach this point in our model. This is 
because the 
stereographic projection we use in the quantum case gives the trefoils 
\[ |v\pm 
\frac{\l^{3}}{2}|^{2}=|u^{3/2}\mp \frac{\l^{3}}{2}|^{2}=\frac{2R^{6}-\l^{6}}{4}, \]
subject to $|u|^{3}+|v|^{2}=R^{6}$.

Another interesting direction of further study would be to look at the case
of SU(3) with 3 flavours of fundamental hypermultiplet. There is evidence that 
at certain points of this theory there is a correspondence to a particular 
2-dimensional (2,2)-supersymmetric U(1) gauge theory with 3 flavours of 
twisted chiral multiplets \cite{hh, nick1, dnt}. This two dimensional theory 
has a strong-coupling 
expansion from which one can extract results about the four dimensional theory
unobtainable by other means. Agreement with this correspondence would lend 
further weight to the veracity of both ideas.

\section*{Acknowledgements}
We would like to thank Nick Dorey, Jerome Gauntlett, Tim Hollowood, 
Stephen Howes, David Olive, Peter West, and especially Dave Tong for useful 
conversations and suggestions. 

\end{document}